\documentclass[%
 reprint,
 amsmath,amssymb,
 aps,
pra,
]{revtex4-1}

\usepackage{graphicx}
\usepackage{dcolumn}
\usepackage{bm}

\RequirePackage{tikz}

\newcommand{\paren}[1]{\left({#1}\right)}
\newcommand{\sqpr}[1]{\left[{#1}\right]}
\newcommand{\dydx}[2]{\frac{\mathrm{d} #1}{\mathrm{d} #2}}

\newcommand{\sech}{\mathrm{sech}}

\newcommand{\dint}{\mathrm{d}}
\newcommand{\dt}{\mathrm{d}t}
\newcommand{\pypx}[2]{\frac{\partial #1}{\partial #2}}
\newcommand{\pynpx}[3]{\frac{\partial^{#3} #1}{\partial {#2}^{#3}}}

\renewcommand{\vec}[1]{\mathbf{#1}}

\renewcommand{\frac}[2]{\dfrac{#1}{#2}}

\newcommand{\bmat}[1]{\begin{bmatrix}#1\end{bmatrix}}

\newcommand{\abs}[1]{\left|{#1}\right|}
\newcommand{\vm}{\mathsf{V}_0}
\newcommand{\wm}{\mathsf{W}_0}
\newcommand{\quoeq}[1]{Eq.~(\ref{#1})}

\begin{document}

\preprint{APS/123-QED}

\title{Nonlinear Stabilization of High-Energy and Ultrashort Pulses\\ in Passively Modelocked Lasers with Fast Saturable Absorption}

\author{Shaokang Wang}
 \email{swan1@umbc.edu}
\author{Brian S. Marks}
\author{Curtis R. Menyuk}%
\affiliation{%
Department of Computer Science and Electrical Engineering, \\
University of Maryland, Baltimore County,  1000 Hilltop Circle, Baltimore, MD, 21250
}%

\date{\today}

\begin{abstract}
The two most commonly used models for passively modelocked lasers with fast saturable absorbers are the Haus modelocking equation (HME) and the cubic-quintic modelocking equation (CQME). 
The HME corresponds to a special limit of the CQME in which only a cubic nonlinearity in the fast saturable absorber is kept in the model. 
Here, we use singular perturbation theory to demonstrate that the CQME has a stable high-energy solution for an arbitrarily small but non-zero quintic contribution to the fast saturable absorber. 
As a consequence, we find that the CQME predicts the existence of stable modelocked pulses when the cubic nonlinearity is orders of magnitude larger than the value at which the HME predicts that modelocked pulses become unstable. 
This intrinsically larger stability range is consistent with experiments. 
Our results suggest a possible path to obtain high-energy and ultrashort pulses by fine tuning the higher-order nonlinear terms in the fast saturable absorber. 
\end{abstract}

\maketitle


\section{Introduction\label{sec:intro}}

Over the past few decades, ultrashort optical pulses that are produced by passively modelocked lasers have been used in many fields~\cite{Diddams:10,kartner2014few}.  
The most commonly used model of passively modelocked lasers with a fast saturable absorber and a slow saturable gain is the Haus modelocking equation (HME)~\cite{haus902165}. 
The original HME may be written as~\cite{haus:3049}
\begin{equation}\label{eq:hme}
\begin{aligned}\displaystyle
\pypx{u}{z}=\bigg[-\frac{l}{2}+\frac{g(\abs{u})}{2} & \bigg(1+\frac{1}{2\omega_g^2}\pynpx{}{t}{2} \bigg) - \frac{i\beta''}{2}\pynpx{}{t}{2}  \\
& + i\gamma|u|^2 -i\phi \bigg]u + f_{\mathrm{sa}}(|u|)u, 
\end{aligned}
\end{equation}
where $u$ is the complex field envelope, $t$ is the retarded time, $z$ is the propagation distance, $\phi$ is the phase rotation per unit length, $l$ is the linear loss coefficient, $g(|u|)$ is the saturated gain, $\beta''$ is the group velocity dispersion coefficient, $\gamma$ is the Kerr coefficient, $\omega_g$ is the gain bandwidth, and $f_{\mathrm{sa}}(|u|)$ is the fast saturable absorption. 
In this article, we focus on the case in which the chromatic dispersion is anomalous with $\beta''<0$.
In the HME, it is assumed that the gain response of the medium is much longer than the roundtrip time $T_R$, in which case the saturable gain becomes
\begin{align}\label{eq:gain_sat}
g(\abs{u})=g_0/[1+P_\mathrm{av}(\abs{u})/P_\mathrm{sat}],
\end{align}
where $g_0$ is the unsaturated gain, $P_{\mathrm{av}}(\abs{u})$ is the average power, 
and $P_\mathrm{sat}$ is the saturation power. 
We may write $P_\mathrm{av}(\abs{u})=\int_{-T_R/2}^{T_R/2} |u(t,z)|^2 \dt/T_R$.
Compared to the pulse duration, the saturable absorption is instantaneous and can be modeled by a nonlinear gain term, 
\begin{align}\label{eq:fsa-cubic}
f_{\mathrm{sa}}(|u|)=\delta|u|^2.
\end{align}
Due to the multiplication of $f_\mathrm{sa}$ and $u$ in Eq.~(\ref{eq:hme}), we refer to $\delta$ as the cubic coefficient of the fast saturable absorption. 

The HME has been successfully used to explore many of the qualitative features of passively modelocked lasers.
However, the predictions from the HME for the instability thresholds are unrealistically pessimistic~\cite{Kapitula:02,Chong:06,Cundiff2003}. 
One particular reason is that when the cubic coefficient $\delta$ becomes sufficiently large,  Eq.~(\ref{eq:fsa-cubic}) provides unlimited gain that leads to an infinite growth of the pulse energy as it propagates~\cite{Kapitula:02}. 
In order to obtain a broader stability regime in modelocked laser models that better matches the experimental observations, the HME has been extended by replacing Eq.~(\ref{eq:fsa-cubic}) with different models of fast saturable absorption.
A common replacement is the cubic-quintic model,
\begin{align}\label{eq:cubic-quintic}
f_{\mathrm{sa}}(|u|)=\delta |u|^2-\sigma |u|^4,
\end{align}
where $\sigma$ is the quintic coefficient that provides a higher-order saturation to the unlimited third-order nonlinear gain~\cite{Soto-Crespo:96, akhmediev10400829,newbury2005jqe}. 

Another approach to modeling modelocked lasers is to treat the gain response as instantaneous, just like the loss, so that the gain saturation is absorbed into the loss saturation. 
This model of modelocking leads to the complex Ginzburg-Landau equation (CGLE), and the modelocked solutions are referred to as dissipative solitons~\cite{Soto-Crespo:96, akhmediev10400829, renninger2008}. 
Computational studies have shown that this model predicts a large stable region in the parameter space in which modelocked pulses exist. 
These pulses can have energies that are many times larger than the energies at which the HME predicts stable operation, and these pulses can exist in both the normal and anomalous dispersion regime. 
However, no real lasers have an instantaneous gain response, and it is difficult to relate the parameters of this theoretical model to the parameters that can be adjusted in experiments. 
By contrast, slow gain saturation plays a fundamental role in stabilizing modelocked pulses in the HME, as is the case in experimental systems~\cite{Chen:94}. 
Both the HME and the CGLE have analytical solutions that have been described in~\cite{Soto-Crespo:96}. 
However, the analytical solutions of the CGLE are unstable and are unconnected with the stable analytical solutions of the HME. 
Hence, computational methods are needed to find a high-energy parameter regime in which lasers with slow saturable gain can operate stably~\cite{Soto-Crespo:96}. 

A laser model that includes both the slow saturable gain that is present in the HME and higher-order saturable absorption that is present in the CGLE is the cubic-quintic modelocking equation (CQME)
\begin{equation}\label{eq:cqme}
\begin{aligned}\displaystyle
\pypx{u}{z}=&\bigg[-\frac{l}{2}+\frac{g(\abs{u})}{2} \bigg(1+\frac{1}{2\omega_g^2}\pynpx{}{t}{2} \bigg) - \frac{i\beta''}{2}\pynpx{}{t}{2}  \\
& + i\gamma|u|^2 -i\phi \bigg]u + \delta |u|^2u-\sigma |u|^4u,
\end{aligned}
\end{equation}
which is given by substituting~\quoeq{eq:cubic-quintic} into~\quoeq{eq:hme}.
Both phenomena are present in real lasers. 
In this work, we will find the high-energy solutions of the CQME, and we will describe the role that the higher-order nonlinearity plays in stabilizing these solutions. 

In prior work~\cite{Wang:2014}, we computationally found the stable region in the $(\sigma,\delta)$ parameter space for~\quoeq{eq:cqme}, using a parameter set that corresponds to a soliton laser ($\beta''<0$). 
We found that for a range of $\sigma$-values, two stable modelocked pulse solutions exist. 
There is a low-amplitude solution that coincides with the stable solution of the HME when $\sigma\to0$ and is stable over a limited range of $\delta$-values. 
Additionally, there is a high-energy solution that for the values of $\sigma$ that we explored remains stable up to $\delta\approx9.5$, which is about a factor of 280 greater than the HME\rq{}s stability limit. 
This work left open the question of what happens when $\sigma\to0$ and~\quoeq{eq:cqme} becomes identical to the HME. 

In this work, we will investigate in detail the limit of~\quoeq{eq:cqme} when $\sigma\to0$. 
Using singular perturbation theory, we will show that the high-energy solution persists regardless of how small $\sigma$ becomes, as long as it is non-zero. 
We will also show that the energy of this solution increases as $\sigma\to0$, suggesting a path towards obtaining high-energy, ultrashort solutions. 
Since any real modelocked laser system with a fast saturable absorber will have a quintic component, this result also shows that the HME cannot be relied upon to quantitatively determine the stability in real systems. 

In Sec.~\ref{sec:bta}, we briefly review the stability structure of the CQME for the parameter set that we consider. 
In Sec.~\ref{sec:stationary} and Sec.~\ref{sec:stability}, we study the high-energy solution using singular perturbation theory~\cite{barenblatt1996}, which enables us to find the solution to~\quoeq{eq:cqme} when $\sigma\to0$ and to determine its stability. 
The energy of this solution increases and its duration decreases as $\sigma\to0$, but the range of $\delta$-values in which it is stable does not change significantly. 
In Sec.~\ref{sec:discussion}, we discuss how these solutions could be obtained experimentally.

\section{stability of the cubic-quintic modelocking equation\label{sec:bta}}

There are two physics-based models of fast saturable absorption from which the HME or the cubic-quintic model is derived.
The first and oldest of these models is due to Haus~\cite{haus902165,haus:3049}. In this model, it is assumed that the absorbing medium is a two-level system in which the response time of the medium is fast compared to the pulse duration, so that the population of the upper state is proportional to $|u(t)|^2$. 
In this case, we find that~\cite{haus902165,haus:3049}
\begin{align}\label{eq:two-level}
\pypx{u}{z}\bigg|_\mathrm{ab}=f_\mathrm{ab}(|u|)u = -\frac{f_0 u}{1+|u(t)|^2/P_\mathrm{ab}},
\end{align}
where $\partial u/\partial z|_\mathrm{ab}$ is the contribution to the loss from the absorbing mechanism, $f_0$ is a constant, and $P_\mathrm{ab}$ is the saturation power of the absorber. 
If $|u(t)|_\mathrm{ab}^2\ll P_\mathrm{ab}$, then we find
\begin{align}\label{eq:solid-taylor}
f_\mathrm{ab}(|u|) = -f_0+\frac{f_0}{P_\mathrm{ab}}|u(t)|^2-\frac{f_0}{P^2_\mathrm{ab}}|u(t)|^4+\cdots.
\end{align}
If we truncate this expansion at the order $|u|^4$, we find that $l/2=\alpha+f_0$, where $\alpha$ donates the total loss that is not due to the material absorber, such as losses from the end mirrors and couplers. 
We also find $\delta=f_0/P_\mathrm{ab}$ and $\sigma=f_0/P_\mathrm{ab}^2$. 

The second physics-based model, due to Chen et al.~\cite{Chen:92}, assumes that saturable absorption is due to a combination of nonlinear polarization rotation and polarization selective elements that attenuate low intensities more than high intensities. 
In this model, we find that~\cite{Chen:92,leblond2002, komarov2005, kamarov-quintic2005}
\begin{align}
\pypx{u}{z}\bigg|_\mathrm{ab}=f_\mathrm{ab}(|u|)u=-f_0+f_1\cos\paren{\mu |u|^2-\nu},
\end{align}
where the constants $f_0$, $f_1$, $\mu$, and $\nu$ depend on the settings of the polarization selective elements and the amount of nonlinear polarization rotation in one pass through the laser. 
If we may assume $\mu|u|^2\ll1$, then
\begin{equation}
\begin{aligned}\label{eq:expand-sinusoidal}
f_\mathrm{ab}(|u|)u=&-f_0+f_1\cos\nu+\mu f_1(\sin\nu) |u|^2 \\
& - (\mu^2f_1/2)(\cos\nu)|u|^4-\cdots.
\end{aligned}
\end{equation}
The combination $-f_0+f_1\cos\nu$ may be absorbed into the total linear loss, and we find $\delta=\mu f_1\sin\nu$ and $\sigma=(\mu^2 f_1/2)\cos\nu$. 

It is useful at this point to normalize~\quoeq{eq:cqme}~\cite{Kapitula:02,kutz:06}. 
We normalize $u(t)$ with respect to the peak amplitude of the electric field $U_0$, the propagation variable $z$ is normalized with respect to a characteristic dispersion length $z_0$, and the retarded time $t$ is normalized with respect to a characteristic pulse time $t_0$. 
Letting $u_n=u/U_0$, $z_n=z/z_0$, and $t_n=t/t_0$, Eq.~(\ref{eq:cqme}) becomes
\begin{equation}
\begin{aligned}
\pypx{u_n}{z_n}=&\bigg[-i\phi z_0 - \frac{lz_0}{2} + \frac{g(|u|)z_0}{2}\paren{1+\frac{1}{2 \omega_g^2 t^2_0}\pynpx{}{t^2_n}{}} \\
&- \frac{i\beta''z_0}{2t_0^2}\pynpx{}{t_n^2}{}+i\gamma z_0 U_0^2 |u_n|^2\bigg] u_n \\
& + \delta z_0 U_0^2 |u_n|^2 u_n - \sigma z_0 U_0^4 |u_n|^4 u_n.
\end{aligned}
\end{equation}
Defining normalized parameters --- $\phi_n=\phi z_0$, $l_n=l z_0$, $g_n(|u|)=g(|u|)z_0$, $\omega_{gn}=\omega_g t_0$, $\beta''_n=\beta'' z_0/t_0^2$, $\gamma_n=\gamma z_0 U_0^2$, $\delta_n = \delta z_0 U_0^2$, and $\sigma_n = \sigma z_0 U_0^4$ --- we obtain a normalized version of~\quoeq{eq:hme}. 
From hereon, we will use these normalized parameters, and we drop the subscript ``$n$.''  
We show the set of normalized parameters that we use in Table~\ref{tab:params}. 
These values are the same as in Table.~1 of~\cite{Wang:2014}; the value $\omega_g=\sqrt{10}/2$ reported there is an error.

\begin{table}[!h]
\begin{center}
\begin{tabular}{|l|c|c|c|c|c|c|}
\hline 
Parameter& $g_0$ & $l$ & $\gamma$ & $\omega_g$ &  $\beta''$ & $T_RP_\mathrm{sat}$ \\
\hline
Value & $0.4$ & 0.2 & $4$ & $\sqrt{5}$ & $-2$  & 1 \\
\hline
\end{tabular}
\end{center}
\vspace{-0.3cm}
\caption{Normalized values of parameters.\label{tab:params}}
\vspace{-0.3cm}
\end{table}

\begin{figure}[h!]
\begin{center}
\includegraphics[scale=1]{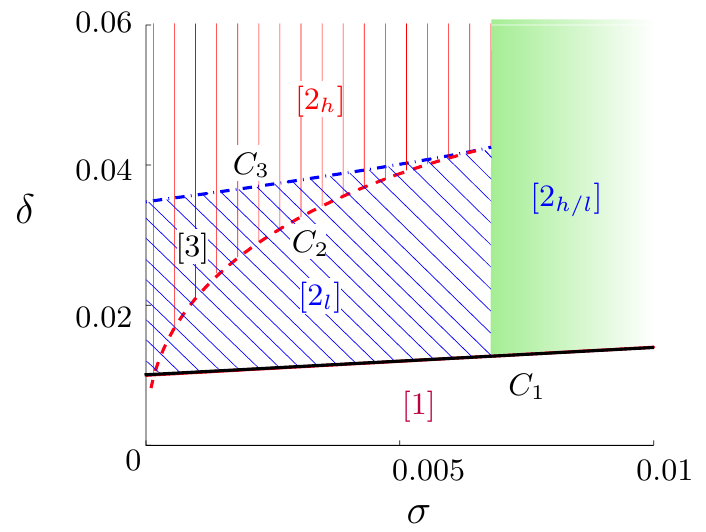} 
\end{center}
\vspace{-0.5cm}
\caption{The stability regions of the CQME with a cubic-quintic saturable absorber $f_\mathrm{sa,cq}(|u|)$. The stability boundaries are marked by three curves, $C_1$, $C_2$, and $C_3$. This figure is reproduced from Fig.~16 of~\cite{Wang:2014}. \label{fig:stability-cqme}}
\end{figure}

We previously used boundary tracking algorithms~\cite{Wang:2014,Wang_IPC2013} to find the regions in the $(\sigma,\delta)$ parameter plane in which modelocked pulses are stable. 
We show the results in Fig.~\ref{fig:stability-cqme}.
The regions of the stability are bounded by three curves $C_1$, $C_2$, and $C_3$. 
On each of these curves, a modelocked pulse solution of the CQME becomes unstable. 
The $\delta$-axis of Fig.~\ref{fig:stability-cqme} corresponds to the HME, since $\sigma=0$. 
In this case, the HME has the analytical solution
\begin{align}\label{eq:hmesolution}
u_h(t) = A_h\sech^{(1+i\beta_h)}\paren{t/t_h},
\end{align}
where $A_h$, $\beta_h$, and $t_h$ are constants that depend on the system parameters.
The stationary solution $u_h(t)$ is stable when $0.01<\delta<0.0348$~\cite{Kapitula:02}.

When $\sigma\ne0$, Eq.~(\ref{eq:cqme}) does not have analytical pulse solutions for the parameter set in Table.~\ref{tab:params}, and the modelocked pulse solutions must be found computationally.
In the region in Fig.~\ref{fig:stability-cqme} that is denoted $[1]$, we have found that there are no stable modelocked pulse solutions. 
In the regions denoted $[2_l]$ and $[3]$, we have found that a modelocked pulse solution exists that we refer to as the low-amplitude solution (LAS) and which is a continuation of the solution of the HME when $\sigma=0$. 
In region $[3]$, there is an additional high-amplitude solution (HAS) that no longer exists when $\delta$ becomes sufficiently small. 
This solution remains stable in the region denoted $[2_h]$, which extends to much higher values of $\delta$ than are shown in Fig.~\ref{fig:stability-cqme}. 
We have found that the HAS becomes unstable when $\delta\approx9.5$ with a non-zero $\sigma$. 
In  Fig.~\ref{fig:stability-cqme}, the curve $C_1$ indicates points along which the LAS becomes unstable; 
the curve $C_2$ indicates points along which the HAS becomes unstable; 
the curve $C_3$ indicates points along which the LAS becomes unstable. 
In the region denoted $[2_{h/l}]$, there are no longer distinct low-amplitude and high-amplitude solutions: there is just one stable solution. 

We previously found computationally that when $\delta$ is as large as 9.5, the HAS remains stable when $\sigma$ is as small as $7\times 10^{-4}$~\cite{Wang:2014}. 
However, this study did not determine what happens to the HAS for a physically reasonable range of $\delta$ as $\sigma\to0$. 
It was unclear whether a stable HAS continues to exist, becomes unstable, or disappears. 
In Secs.~\ref{sec:stationary} and~\ref{sec:stability}, we will show that the HAS exists and remains stable for any non-zero $\sigma$ as long as $\delta\lesssim9.5$. 
At the same time, its energy increases and its duration decreases --- opening up a potential path to obtain high-energy pulses.

\section{The stationary pulse as $\sigma\to0$\label{sec:stationary}}

\begin{figure}[!h]
\begin{center}
\vspace{0.1cm}
\includegraphics[scale=1]{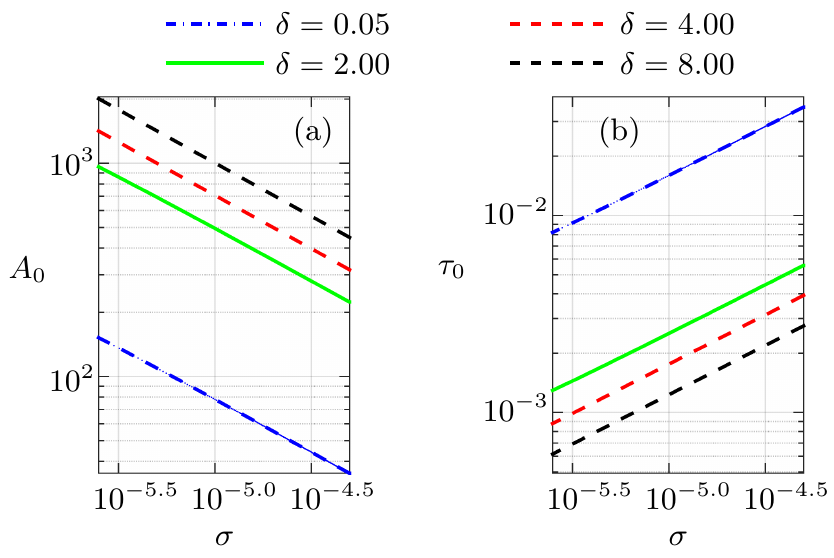}
\end{center}
\vspace{-0.5cm}
\caption{(a) The peak amplitude $A_0$ and (b) the FWHM pulse duration $\tau_0$ of the computational stationary pulse solution of the CQME as $\sigma\to0$ and $\delta$ varies. 
The slopes of the curves equal $-1/2$ and $1/2$ in (a) and (b) respectively.
\label{fig:amplitude}}
\vspace{-0.2cm}
\end{figure}

The computational approach that we used to obtain Fig.~\ref{fig:stability-cqme} does not continue to work well for the HAS when $\sigma\to0$ because the pulse becomes singular; its energy increases and its duration decreases. 
This behavior is visible in Fig.~\ref{fig:amplitude}, where we see the variation of the peak amplitude $A_0$ and its FWHM duration as $\sigma$ decreases. 
An alternative approach is therefore required to determine whether a modelocked pulse exists in this limit and---if it continues to exist---whether it is stable. 
We use singular perturbation theory to address these questions. 

\subsection{The Dominant Balance}

From Fig.~\ref{fig:amplitude}, we infer for all values of $\delta$ that $A_0\propto\sigma^{-1/2}$ as $\sigma\to0$. 
Based on this observation, we seek a stationary (equilibrium) solution of~\quoeq{eq:cqme} that has the form 
\begin{equation}
\begin{aligned}\label{eq:insigma}
\phi_0=\psi_0\sigma^{-1}, \quad u_0(t)=\sigma^{-1/2}a_0(\sigma^{-1/2}t),
\end{aligned}
\end{equation}
For the stationary solution, we must have $\dint a_{0}/\dint z=\dint \psi_0/\dint z=0$.
We will find that the equations that govern $a_0$ and $\psi_0$ become independent of $\sigma$ in the limit $\sigma\to0$, which allows us to determine them. 

We let $\tau = \sigma^{-1/2}t$, and we use a prime to denote derivatives with respect to $\tau$, so that
\begin{align}
\pynpx{u_0}{t}{} = a_0'\sigma^{-1}, \quad \pynpx{u_0}{t}{2} = a_0''\sigma^{-3/2}.
\end{align}
We also find
\begin{align}\label{eq:gain_sat_2}
g(|u|) = g_0/(1+C_g\sigma^{-1/2}),
\end{align}
where $C_g=\int_{-\infty}^{\infty}|a_0(\tau)|^2\dint \tau/(P_\mathrm{sat}T_R)$. 
After substitution of Eq.~(\ref{eq:gain_sat_2}) into Eqs.~(\ref{eq:cqme}) we find
\begin{equation}
\begin{aligned} \label{eq:expansion_sigma2}
\paren{C_g\sigma+\sigma^{3/2}} \pynpx{a_0}{z}{}& =\dfrac{g_0-l}{2}a_0\sigma^{3/2} -\dfrac{l}{2}C_g a_0\sigma \\ & +\sigma^{1/2}\sqpr{\dfrac{g_0}{4\omega_g^2}a''_0 + f} + C_g f, \quad
\end{aligned}
\end{equation}
where $f=\paren{\delta+i\gamma}|a_0|^2a_0- |a_0|^4a_0  - i\psi a_0 -{i\beta'' a_0''}/{2}$. 
As $\sigma\to0$ and when $a(\tau)\ne0$, the dominant balance of this system is
\begin{align}\label{eq:dominant-balance}
f=\paren{\delta+i\gamma}|a_0|^2a_0-|a_0|^4a_0 - i\psi_0 a_0-\dfrac{i\beta''}{2}a_0'' = 0,
\end{align}
from which we solve for the asymptotic stationary solution $[a_0(\tau),\psi_0]$.

The balance of the dominant terms in Eq.~(\ref{eq:dominant-balance}) implies that, in the CQME of Eq.~(\ref{eq:cqme}), as $\sigma\to0$, the gain and the loss are balanced via the cubic term $\delta|a_0|^2a_0$ and the quintic term $\sigma|a_0|^4a_0$, while the saturated gain and the linear loss play no role in forming the stationary pulse. 
The remaining imaginary terms imply that the pulse envelope $a_0(\tau)$ is in general complex, i.e., a chirp is required to satisfy $f=0$ in \quoeq{eq:dominant-balance}. 

\subsection{The Asymptotic Stationary Pulse}

\begin{figure}[!h]
\begin{center}
\includegraphics[scale=1]{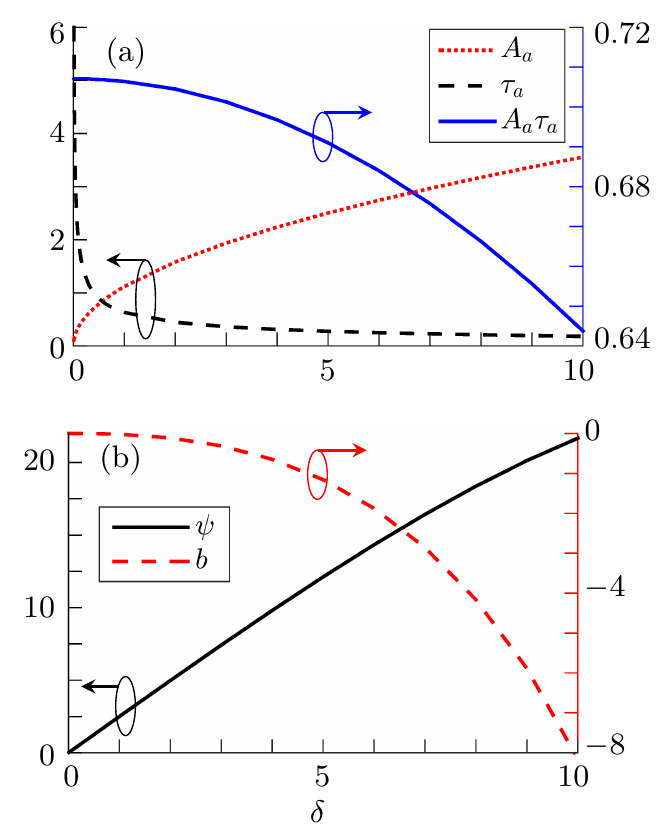}
\end{center}
\vspace{-0.5cm}
  \caption{(a) The peak amplitude $A_a$, the pulse-width $\tau_a$, and their product $A_a\tau_a$, (b) the rotation rate coefficient $\psi$, and the quadratic chirp coefficient $b$ of the asymptotic stationary solution that is obtained by finding the root of $f$ in Eq.~(\ref{eq:dominant-balance}). \label{fig:a&theta}}
\end{figure}

We use the nonlinear root-finding method that is described in~\cite{Wang:2014} to computationally solve Eq.~(\ref{eq:dominant-balance}). 
We consider the parameter set that is shown in Table.~\ref{tab:params}.
In Fig.~\ref{fig:a&theta}, we show the profile of the asymptotic solution that we have found computationally, in which $A_a$ is the peak amplitude of $a_0(\tau)$, $\tau_a=\tau_{a,\mathrm{FWHM}}/0.57$, where $\tau_{a,\mathrm{FWHM}}$ is the FWHM width of $a_0(\tau)$, and the chirp coefficient is given by
\begin{align}
b=\mathrm{Im}\frac{\int_{-\infty}^{\infty}\tau a_0^*a_0'\dint \tau}{\int_{-\infty}^{\infty}\tau^2|a_0|^2\dint \tau}.
\end{align}
 As $\delta$ increases, the amplitude $A_a$ increases while $\tau_a$ decreases, i.e., the asymptotic stationary pulse solution becomes increasingly taller and narrower. 
Meanwhile, we find that $A_a\tau_a\approx\sqrt{2}$ when $\delta\approx0$, and decreases as $\delta$ grows. 
Hence the pulse shape is close to that of a nonlinear Schr\"odinger (NLS) equation soliton when the nonlinear gain is small, and it deviates from the NLS soliton profile as the nonlinear gain grows. 
In addition, when $\delta\approx0$, the phase rotation rate coefficient $\psi$ is close to 0, while the pulse is almost chirp-free. 
Then, as $\delta$ increases, we find that $\psi$ increases, and the chirp across the pulse increases.

\begin{figure}[!h]
\begin{center}
\includegraphics[scale=1]{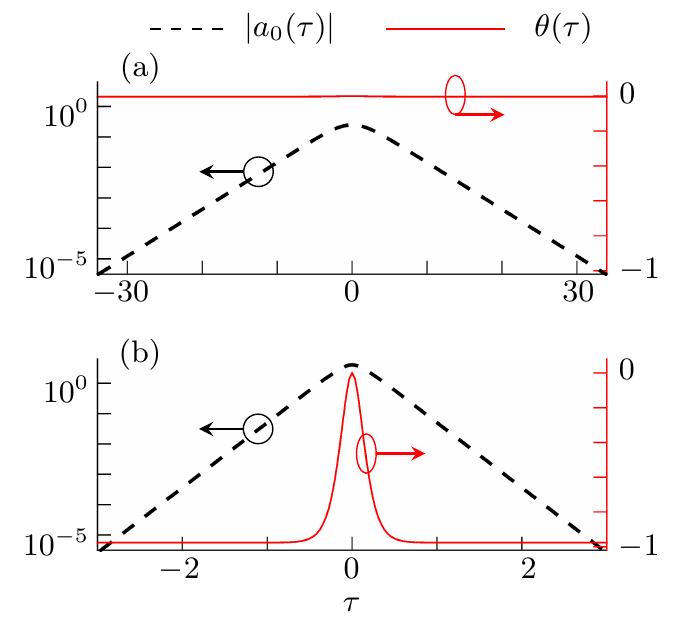}
\end{center}
\vspace{-0.5cm}  
  \caption{The asymptotic stationary solution obtained by solving Eq.~(\ref{eq:dominant-balance}) with (a) $\delta=0.05$ and (b) $\delta=13.00$. Here, $\theta(\tau)$ is the phase change across the pulse in radians. 
Note that the scales of $\tau$ are different in the two sub-figures.    \label{fig:asymptoticpulse}}
\end{figure}

The amplitude of the asymptotic pulse solution that we have found is similar in shape to a hyperbolic-secant pulse, in which the wings of the pulse decay exponentially as $|t|$ increases. 
We show two examples of asymptotic pulses with $\delta=0.05$ and $\delta=13.00$ in Fig.~\ref{fig:asymptoticpulse}, in which $\theta(\tau)$ is the phase change across the pulse, i.e., $a_0(\tau)=|a_0(\tau)|\exp[i\theta(\tau)]$. 
The variation of $\theta(\tau)$ increases significantly as $\delta$ increases, which is consistent with the change in the chirp parameter $b$ that is shown in Fig.~\ref{fig:a&theta}. 

Afanasjev~\cite{Afanasjev:95} has reported that the analytical pulse solutions of the CGLE becomes singular when both the linear gain and the quintic coefficient vanish, which is similar to our result. 
However, these analytical solutions are always unstable and cannot be used to model modelocked lasers~\cite{Soto-Crespo:96,Wang:2014,renninger2008}. 

\section{Stability of the CQME as $\sigma\to0$\label{sec:stability}}

Next, we evaluate the stability of these stationary pulse solutions.
We first linearize Eq.~(\ref{eq:expansion_sigma2}) about the stationary solution, and we determine the spectrum (eigenvalues) of this linearized equation. 
The spectrum that we find in this case is similar to the spectrum that appears in soliton perturbation theory~\cite{haus206583,kaup1990}.
There are two branches of eigenvalues that correspond to continuous wave perturbations, and there are four discrete modes that correspond to perturbations of the stationary solution's central time, central phase, amplitude, and central frequency, and whose eigenvalues we will denote as $\lambda_\tau$, $\lambda_\phi$, $\lambda_a$, and $\lambda_f$ respectively. 
The solution is linearly stable if the real part of the two continuous branches are negative and the discrete eigenvalues $\lambda_f$ and $\lambda_a$ are both negative, while $\lambda_\tau$ and $\lambda_\phi$ remain at the origin due to time and phase invariance of Eq.~(\ref{eq:cqme}). 

\subsection{Linearization}

When we linearize Eq.~(\ref{eq:expansion_sigma2}), we can neglect the terms that are proportional to $\sigma^{3/2}$, as these terms tend to zero faster than terms proportional to $\sigma^m$ with $m<3/2$ as $\sigma\to0$. 
If we add a perturbation $\Delta a$ to the stationary pulse solution $a_0(\tau)$, and then linearize Eq.~(\ref{eq:expansion_sigma2}) about $a_0(\tau)$, we then obtain
\begin{align} \label{eq:linearization1}
\sigma\pypx{\Delta a}{z}  \approx
 -\dfrac{l}{2}\sigma \Delta a+ \frac{\sigma^{1/2}}{C_g }\paren{f_a+\dfrac{g_0}{4\omega_g^2}\Delta a''}+f_a,
\quad
\end{align}
where $f_a$ is the derivative of $f$ with respect to $\Delta a$,
\begin{equation}
\begin{aligned}\label{eq:nonlinear-terms}
f_a=& \paren{\delta+i\gamma} \paren{2|a_0|^2\Delta a+a_0^2\Delta a^*} -{i\beta''}/{2}\ \Delta a''  \\
&- i\psi_0 \Delta a - 3|a_0|^4\Delta a - 2|a_0|^2a_0^2\Delta a^*.
\end{aligned}
\end{equation}

\subsection{Continuous Waves}

The stability condition for the continuous modes is $g(|u|)-l<0$~\cite{Wang:2014}.
This condition becomes $-l<0$ in the limit $\sigma\to0$ since the pulse energy grows exponentially and thus $g(|u|)\to0$.
This behavior appears in our asymptotic solution. 
As illustrated in Fig.~\ref{fig:asymptoticpulse}, the pulse envelope $|a_0(\tau)|$ decays exponentially as $|\tau|\to\infty$, with a decay rate that becomes infinite as $\sigma\to0$. 
As a consequence, the terms proportional to $|a_0|^2$ and $|a_0|^4$ in Eq.~(\ref{eq:nonlinear-terms}) become negligible, and Eq.~(\ref{eq:linearization1}) becomes
\begin{equation}
\begin{aligned}\label{eq:cont-linearization}
\sigma\pypx{\Delta a}{z}=&-\frac{l}{2}\sigma\Delta a + \frac{\sigma^{1/2}}{C_g }\dfrac{g_0}{4\omega_g^2}\Delta a''\\
& -  i\paren{1+\frac{\sigma^{1/2}}{C_g }}\paren{ \frac{\beta''}{2} \Delta a'' + \psi_0 \Delta a}.
\end{aligned}
\end{equation}
In the Fourier domain, Equation~(\ref{eq:cont-linearization}) becomes
\begin{align}
\pypx{\Delta \tilde{a}}{z}=\lambda_c(\omega)\Delta\tilde{a}
\end{align}
where $\Delta \tilde{a}(\omega)$ is the Fourier transform of $\Delta a(\tau)$ and
\begin{align}\label{eq:stab-cond-cont}
\mathrm{Re}\{\lambda_c(\omega)\} = -\paren{\frac{l}{2}+\frac{g_0}{4C_g\sigma^{1/2}\omega_g^2}\omega^2}.
\end{align}
Equation~(\ref{eq:stab-cond-cont}) implies that the stationary pulse solution is always stable with respect to continuous modes with $l>0$, a result that agrees with our previous conclusion in~\cite{Wang:2014}. 

\subsection{Discrete Modes}

The discrete modes can be evaluated computationally by performing an eigenanalysis of the Jacobian of Eq.~(\ref{eq:linearization1}). 
Here, we study the case when $a_0\ne0$ and $\sigma\to 0$. The stability of $\Delta a$ will be dominated by the zero-order terms in powers of $\sigma$ on the right hand side of Eq.~(\ref{eq:linearization1}), so that
\begin{align}\label{eq:linearized-discrete}
\sigma \pypx{\Delta a}{z}=f_a,
\end{align}
where $f_a$ is defined in Eq.~(\ref{eq:nonlinear-terms}).
We can then determine the stability of the asymptotic stationary solution by analyzing the spectrum of the Jacobian of the system that is given by Eq.~(\ref{eq:linearized-discrete}).  Because $\Delta a^*$ appears in $f_a$, we must extend Eq.~(\ref{eq:linearized-discrete}) to include the equation for $\partial \Delta a^*/\partial z$ in order to have a complete eigensystem [11], analogous to what is done in soliton perturbation theory.
Instead of directly solving for $\partial \Delta a/ \partial z$ and $\partial \Delta a/ \partial z^*$, it is computationally convenient to let $a_0(\tau) = v_0(\tau) + iw_0(\tau)$.  
We then use $\Delta v = (\Delta a + \Delta a^*)/2$ and $\Delta w = (\Delta a - \Delta a^*)/(2i)$ to denote the perturbations to $v_0$ and $w_0$.  
Similar to~\cite{Wang:2014}, we discretize the system in a computational window $\tau\in[-T_\tau/2, T_\tau/2]$ --- where $a_0(\pm T_\tau/2)\approx0$ --- using $N$ equispaced points $\{\tau=\tau_j,\ j=1,2\ldots,N\}$.
Using Eq.~(\ref{eq:linearized-discrete}), we formulate the extended system and then a linear eigenvalue problem as
\begin{align}
\dydx{}{z}\left[\begin{array}{ccc}\Delta \vec{v}\\ \Delta \vec{w}\end{array}\right]=
{\mathsf J}\left[\begin{array}{ccc}\Delta \vec{v}\\ \Delta \vec{w}\end{array}\right] = 
\lambda\left[\begin{array}{ccc}\Delta \vec{v}\\ \Delta \vec{w}\end{array}\right],
\label{eq:eigen_num}
\end{align}
where the vectors $\Delta\vec{v}$ and $\Delta\vec{w}$ are defined as $\Delta\vec{v}_j = \Delta v(\tau_j)$, and $\Delta\vec{w}_j=\Delta w(\tau_j)$, and the Jacobian matrix $\mathsf{J}$ is
\begin{align}
\mathsf{J}= \bmat{
\mathsf{J}_{11} & \mathsf{J}_{12} \\ \mathsf{J}_{21} & \mathsf{J}_{22} 
},  \label{eq:JG}
\end{align}
where the sub-matrices are defined as
\begin{widetext}
\begin{equation}
\begin{aligned}
\mathsf{J}_{11}&=\delta\paren{3\vm^2+\wm^2}-2\gamma \vm\wm- 5\vm^4 - \wm^4-6\vm^2\wm^2,  \\
\mathsf{J}_{12}&=4\vm\wm\paren{\frac{\delta}{2}-\vm^2-\wm^2}-\gamma\paren{\vm^2+3\wm^2} +\psi+\frac{\beta''}{2}\mathsf{D}^2_\tau,\\
\mathsf{J}_{21}&=4\vm\wm\paren{\frac{\delta}{2}-\vm^2-\wm^2}+\gamma\paren{3\vm^2+\wm^2} -\psi-\frac{\beta''}{2}\mathsf{D}^2_\tau,\\
\mathsf{J}_{22}&=\delta\paren{\vm^2+3\wm^2}+2\gamma\vm\wm  - \vm^4 -5 \wm^4-6\vm^2\wm^2,  
\end{aligned}
\end{equation}
\end{widetext}
in which $\mathsf{D}_\tau^2$ is the second-order differentiation matrix in $\tau$ that is defined in Sec.~4.B.3.~in~\cite{Wang:2014}, and both $\vm$ and $\wm$ are diagonal matrices with 
$\mathsf{V}_{0,jj}=v_0(\tau_j)$, and $\mathsf{W}_{0,jj}=w_0(\tau_j)$. 

\begin{figure}[!h]
\begin{center}
\includegraphics[scale=1]{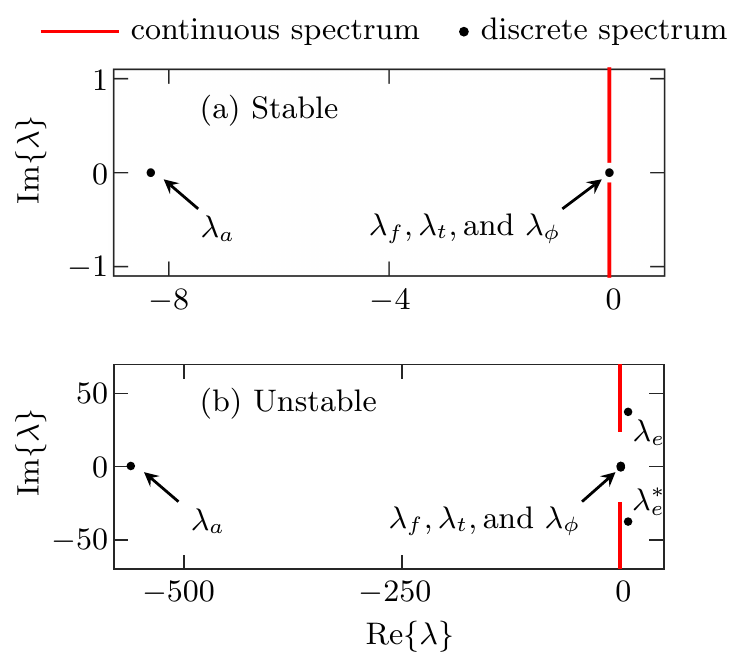}
\end{center}
\vspace{-0.5cm}
\caption{The spectrum of the Jacobian ${\mathsf J}$ in Eq.~(\ref{eq:JG}) with (a)~$\delta=0.05$ and (b)~$\delta=13.00$.  \label{fig:SolitonSpectrum}}
\vspace{-0.2cm}
\end{figure}

We can determine the stability of the asymptotic stationary solution by analyzing the spectrum of the matrix $\mathsf{J}$. 
First, we find the contribution of $f_a$, defined in Eq.~(\ref{eq:nonlinear-terms}), to the stability of the continuous modes by setting $\mathsf{V}_0=\mathsf{W}_0=0$ in Eq.~(\ref{eq:JG}). When evaluated in the frequency domain, we have
\begin{align}
\lambda(\omega) = \pm i |\psi_0-\beta''\omega^2/2|.
\end{align}
The continuous spectrum $\lambda(\omega)$ is purely imaginary, which implies that the dominant balance for \quoeq{eq:expansion_sigma2}, given by $f$ in Eq.~(\ref{eq:dominant-balance}) does not determine the stability; it only indicates the rate of phase rotation of the continuous modes. 
This result does not affect the stability condition for the continuous modes that we described earlier. 
In Fig.~\ref{fig:SolitonSpectrum}(a), we show the spectrum of $\mathsf{J}$ when $\delta=0.05$. 
There are four real discrete eigenvalues, which is similar to the spectrum of the stationary solution the HME.
However, in contrast to the HME, the eigenvalue due to the frequency shift is $0$, which occurs because the dominant balance in Eq.~(\ref{eq:dominant-balance}) corresponds to an unfiltered system --- the frequency filter scales with the saturated gain, which vanishes as $\sigma\to0$. 

In Fig.~\ref{fig:SolitonSpectrum}(b), we show the spectrum of $\mathsf{J}$ when $\delta=13$. 
We observe that an extra pair of discrete eigenvalues, $\lambda_e$ and $\lambda_e^*$, now exist on the positive real side of the complex plane, which implies that the system is unstable at this large value of $\delta$. 
As $\delta$ decreases, the real part of $\lambda_e$ decreases, and both $\lambda_e$ and $\lambda_e^*$ approach and eventually become indistinguishable from the continuous spectrum. 
This result is consistent with the earlier report that a new pair of discrete modes bifurcates from the continuous spectrum when the cubic coefficient $\delta$ grows~\cite{Wang:2014}. 

\begin{figure}[!h]
\begin{center}
\includegraphics[scale=1]{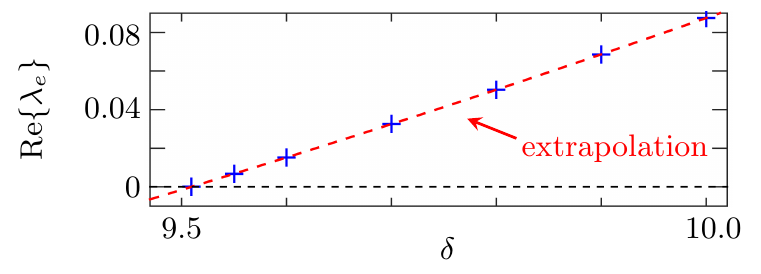}
\end{center}
\vspace{-0.5cm}
\caption{The variation of the real part of the eigenvalue $\lambda_e$ which determines the stability of the asymptotic stationary solution when $\sigma\to0$.  \label{fig:stability}}
\end{figure}

We use the approach that was described in~\cite{Wang:2014} to calculate the eigenvalues as $\delta$ decreases. 
We show the result in Fig.~\ref{fig:stability}. 
We find that the real part of these eigenvalues become 0 at $\delta\approx9.5094$. 
So, the asymptotic stationary solution is stable as long as $\delta<9.5094$, where these two eigenvalues merge into the continuous spectrum and the computation stops.
Compared to the stable range of the HME ($0.01<\delta<0.0348$)~\cite{Kapitula:02}, we find that the stability range of CQME is significantly larger.
This result is consistent with the stability boundary that we have found in cases with small but nonzero values of $\sigma$ in~\cite{Wang:2014}.

\section{Discussion\label{sec:discussion}}

The stable self-similar solution that we have found in this article sheds further light on the dynamical structure of the CQME~\cite{Kapitula:02, Wang:2014}.
We have found in~\cite{Wang:2014} that, in contrast to the HME where there is only one stable solution, two stable equilibrium pulse solution can coexist in a region of the parameter space. 
When $\sigma\to0$, the low amplitude solution tends to the stable solution of the HME, where an analytical expression is available, as long as $\delta$ is below the HME\rq{}s stability limit. 
Here, in this article we prove that the high-amplitude solution remains stable as $\sigma\to0$, although the pulse energy increases and the pulse duration decreases. 
More significantly, our results show that stabilization of the laser system is achieved by a balance between the cubic and the quintic nonlinearity instead of the saturated gain and linear loss, which is the physical reason that the CQME has a much larger region of stability than does the HME even when the quintic term is small. 

In a very large range of $\delta$, when the quintic coefficient disappears, i.e., $\sigma\to0$, the energy of the high-amplitude solution becomes increasingly large. 
This behavior is consistent with the way in which unstable solutions of the HME evolve when $\delta$ is above the instability threshold ($\delta>0.0348$).
The propagating pulse becomes increasingly narrow and energetic, and it eventually blows up.
However, a quintic nonlinearity---no matter how small---is always present in any real laser system, and this quintic nonlinearity will put a halt to the continued growth of the pulse energy. 
This physical insight is consistent with the existence of a large region of stability that has been reported in modelocked lasers~\cite{Cundiff2003}, and this result suggests that the CQME intrinsically provides a better qualitative approximation to practical modelocked lasers than does the HME. 

Our results suggest a possible path toward obtaining high-energy and ultrashort laser pulses. 
The balance of the higher-order nonlinear terms stabilizes these high-energy solutions, so that such solutions can be accessed by decreasing the quintic nonlinearity while keeping the cubic nonlinearity fixed. 
This result can be achieved in principle by adjusting the parameters of the saturable absorber. 
For a laser in which the fast saturable absorber is a two-level system, as described by~\quoeq{eq:two-level}, one would increase the saturable power $P_\mathrm{ab}$ while keeping $f_0/P_\mathrm{ab}$ fixed. 
For a laser that is locked using nonlinear polarization rotation, it would be desirable to set $\sin\nu=0$ in~\quoeq{eq:expand-sinusoidal}, as for example in the configuration of~\cite{Chen:92}. 
This insight may be difficult to apply to real lasers in which the parameters of the saturable absorber lie outside the precise control of experimentalist. 
However, our results demonstrate that there is a strong motivation to better control these parameters. 

\begin{acknowledgments}
{We thank Valentin Besse and Thomas Carruthers for useful comments.  
This work was supported by AMRDEC/DARPA, grant no. W31P4Q-14-1-0002.}%
\end{acknowledgments}

\bibliography{myrefs}		

\end{document}